\documentclass[lettersize,journal]{IEEEtran}
\usepackage{amsmath,amsfonts}
\usepackage{algorithmic}
\usepackage{algorithm}
\usepackage{array}
\usepackage[caption=false,font=normalsize,labelfont=sf,textfont=sf]{subfig}
\usepackage{textcomp}
\usepackage{stfloats}
\usepackage{url}
\usepackage{verbatim}
\usepackage{graphicx}
\usepackage{physics}
\usepackage{cite}
\usepackage{tikz}
\usepackage{circuitikz}
\usepackage{amsmath, amssymb}
\usepackage{moresize} 
\usepackage[electronic]{ifsym} 
\usepackage{bigfoot} 
\usepackage{amssymb}
\usepackage{listings}
\usepackage{cite}
\usepackage{lipsum}
\usepackage{epsfig}
\usepackage{graphicx}
\usepackage{epic}
\usepackage{eepic}
\usepackage{times}
\usepackage{siunitx}
\usepackage{bodegraph}
\usepackage{pgfplots}
\usepackage{adjustbox}
\usepackage{multirow}
\usepackage{color,soul}

\usetikzlibrary{intersections}
\usetikzlibrary{calc}
\usetikzlibrary{positioning}

\definecolor{iron}{rgb}{0.4, 0.6, 0.8}
\definecolor{copper}{rgb}{0.722 0.451 0.2}

\let\OLDthebibliography\thebibliography
\renewcommand\thebibliography[1]{
  \OLDthebibliography{#1}
  \setlength{\parskip}{0pt}
  \setlength{\itemsep}{0pt plus 0.3ex}
}
\usepackage[numbered,framed]{matlab-prettifier}
\tikzset{ext/.pic={
\path [fill=white] (-0.2,0)to[bend left](0,0.1)to[bend right](0.2,0.2)to(0.2,0)to[bend left](0,-0.1)to[bend right](-0.2,-0.2)--cycle;
\draw (-0.2,0)to[bend left](0,0.1)to[bend right](0.2,0.2) (0.2,0)to[bend left](0,-0.1)to[bend right](-0.2,-0.2);
}}

\usepackage[absolute]{textpos}
\usepackage{tabularray}
\UseTblrLibrary{booktabs}

\lstMakeShortInline"

\usetikzlibrary{shapes.geometric, arrows}
\tikzstyle{block} = [rectangle, rounded corners, minimum height=1.5cm,text centered, draw=black, fill=white]
\tikzstyle{block3} = [rectangle, rounded corners, minimum height=1.5cm,text centered, draw=black, fill=white]
\tikzstyle{block1} = [rectangle, rounded corners, text width=0.25*\columnwidth, minimum height=2.5cm,text centered, draw=black, fill=white]
\tikzstyle{block2} = [rectangle, rounded corners, minimum width=0.25cm, minimum height=1cm,text centered, draw=black, fill=white]
\tikzstyle{terminal} = [circle, draw=black, fill=white, very thick, minimum size=0.5mm]
\tikzstyle{conn} = [circle, draw=black, fill=white, very thick, minimum size=0.1mm]
\tikzstyle{arrow} = [very thick,-,>=stealth]
\tikzstyle{arrow1} = [very thick,->,>=stealth]
\tikzstyle{hl} = [rectangle, rounded corners, minimum width=8cm, minimum height=4.5cm,text centered, draw=black, fill=yellow!60, opacity=0.3]
\tikzstyle{arrow2} = [->,>=stealth]
\tikzstyle{arrow3} = [-,>=stealth]
\pgfplotsset{compat=1.18}

\begin{document}

\title{Nonlinear Magnetics Model for Permanent Magnet Synchronous Machines
Capturing Saturation and Temperature Effects}

\author{Kishan Srinivasan,~\IEEEmembership{Student Member,~IEEE,}, Heath Hofmann,~\IEEEmembership{Fellow,~IEEE} and Jing Sun,~\IEEEmembership{Fellow,~IEEE}}




\maketitle

\begin{abstract}
This paper proposes a nonlinear magnetics model for Permanent Magnet Synchronous Machines (PMSMs) that accurately captures the effects of magnetic saturation in
the machine iron and variations in rotor temperature on the permanent magnet excitation. The proposed model considers the permanent magnet as a current source rather than the more commonly used flux-linkage source. A comparison of the two modelling approaches is conducted using Finite Element Analysis (FEA) for different machine designs as well as experimental validation, where it is shown that the proposed model has substantially better accuracy. The proposed model decouples magnetic saturation and rotor temperature effects in the current/flux-linkage relationship, allowing for adaptive estimation of the PM excitation.
\end{abstract}

  
\section{Introduction}
\IEEEPARstart{P}{ermanent} Magnet Synchronous Machines (PMSMs) have become the popular choice in applications such as vehicle electrification~\cite{4168013, 7210190, 5764534, 7112507, 1211236, 178163, 7098414, 4504786, 6064067}. With recent advancements in cooling techniques, it is now possible to develop PMSM designs with very high operating current density, and hence power density~\cite{7862256, 9992038, 6646754}. At these high current excitation levels, PMSMs tend to exhibit significant nonlinear behavior due to saturation of their soft magnetic material. Furthermore, the remanent flux density of NdFeB permanent magnets (PMs) can
decrease significantly with increasing rotor temperature \cite{N35UH}. References \cite{7917257, 7880626, 7873348, 9628538, 370284, 4347951} demonstrate that both magnetic saturation and rotor temperature variation can significantly affect the magnetic properties of the machine, and hence its torque production capability. Thus, it is important to accurately capture these phenomena in machine models used for the purpose of control.

The most common controls model for PMSMs is the linear model, where flux-linkage/current relationship ($\vec{\lambda}^r$ - $\vec{i}^r$) in the rotor reference frame is modelled with an inductance matrix (\textbf{L}) and a permanent magnet flux linkage source ($\Lambda_{PM}$) \cite{9176, 25541, 1064466}. Rotor temperature variation is typically captured by modeling $\Lambda_{PM}$ as a function of rotor temperature $T_r$. These models are inaccurate when the machine iron is magnetically saturated \cite{1064466, 43253}. 
\begin{align}
    \vec{\lambda}{}^r = {\bf L} \vec{i}{}^r + \vec{\lambda}{}^r_{pm} (T_r);
    \end{align}
    \begin{align}
    {\bf L} &= \left[ \begin{array}{cc} L_d & 0 \\ 0 & L_q \end{array} \right], && \vec{\lambda}{}^r = \left[ \begin{array}{c} \lambda_{d}^r \\ \lambda_{q}^r \end{array} \right], && \vec{i}{}^r = \left[ \begin{array}{c} i_{d}^r \\ i_{q}^r \end{array} \right], \nonumber\\
    \vec{\lambda}{}^r_{pm} (T_r) &= \left[ \begin{array}{c} \Lambda_{PM} (T_r) \\ 0 \end{array} \right].
    \end{align}

Many different control-oriented models of PMSMs exist in the literature that attempt to model the nonlinear magnetic properties under certain assumptions \cite{7917257, 7880626, 7873348, 5994848, 43253, 7490387, 111, en12050783, 6408244, 1413524, 1233584, 732255, 7051237, 6349974, 5994804, 4629410, 7001597,7258372}. One of the assumptions that is typically made is to neglect the effects of temperature variation on the soft magnetic material properties, as permanent magnets are much more sensitive to temperature change due to their lower Curie temperature. The most common approach is to model the inductances \cite{5994848,1233584,6349974,5994804,111,en12050783,732255} as a function of  rotor reference frame currents $i_d^r$ and $i_q^r$ to capture the saturation phenomenon, while the permanent magnets (PMs) are  modelled as a flux linkage source whose value varies with rotor temperature.
However, this approach becomes inaccurate in regions of heavy saturation and significant rotor temperature variation \cite{1413524}. A nonlinear model developed using the equivalent magnetic circuit approach is proposed in \cite{4629410}, but does not consider the effects of rotor temperature. Models have also been proposed \cite{7051237,7001597} that take into account saturation and cross saturation by using Look-Up Tables (LUTs). In an
attempt to address issues with accuracy,
\cite{7917257,7880626} propose forming higher-dimensional LUTs with current as a function of both flux linkage and rotor temperature, i.e. $\vec{i}^r$=$f(\vec{\lambda}^r, T_r)$. However, using this approach for real-time control  requires measurement or estimation of the rotor temperature. Another approach is to calculate the inductance matrix as a function of rotor temperature and current from an N-dimensional LUT. A few authors \cite{7873348,7258372} have also suggested to include compensation terms to take into account rotor temperature variation in the LUT, which primarily captures saturation. All these different nonlinear models follow the similar premise of modeling the PM as a flux linkage source. 

The last category of modeling approach involves using Finite Element Analysis (FEA) or reduced-order models formed using FEA \cite{7409111,4676996,8557671}. However, these models are typically computationally intensive and hence not suitable for real-time control purposes. Moreover, obtaining accurate FEA models of actual PMSMs requires knowledge of the dimensional parameters and material properties of the machine, information which is not often available to the controls engineer. 

An alternate approach is to model PM as a current source, as discussed in \cite{18878, Dehkordi2005PermanentMS, 4504786}. However, these are linear magnetic models developed for simulation purposes and not for real-time control applications.
In this paper, we present a magnetics model for PMSMs that accurately captures both the nonlinear and rotor-temperature-dependent behavior of PMSMs by modeling the PM as a current source as opposed the typical flux linkage source. A key contribution of this model is the decoupling of the nonlinear effects of magnetic saturation and the rotor temperature effect on permanent magnet excitation which is captured in a single parameter, the permanent magnet current. This parameter can then be adaptively estimated in real time, removing the need for measurement or estimation of the rotor temperature. The result is an accurate model suitable for use in real-time control.

The proposed PMSM model is introduced and justified in Section \ref{Section2}. A comparison of the proposed model with a common nonlinear model is conducted using FEA simulation results for different machine designs is discussed in \ref{SubSection2}. A similar comparison
is conducted using experimental data in \ref{SubSection4}, followed by concluding remarks in \ref{Section5}. 

\section{Proposed Model}
\label{Section2}
\subsection{Model development}
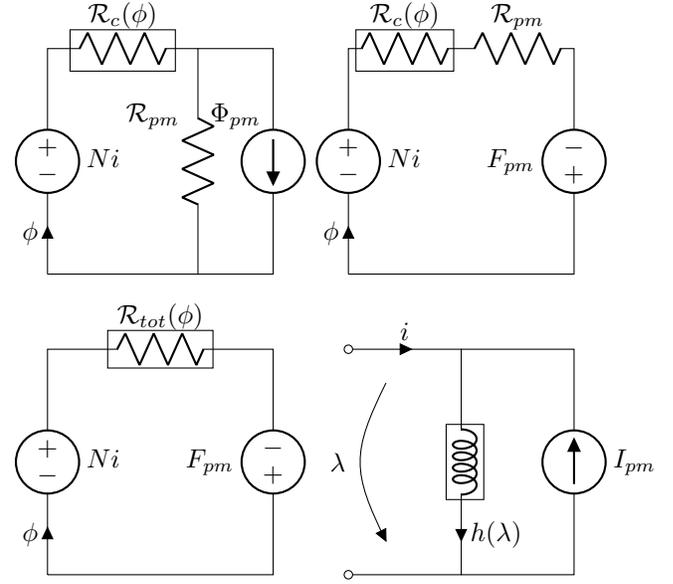
\begin{figure}[htbp]
\centering
\begin{circuitikz}[american voltages]
        \draw (0,0) to [V, invert, a=$Ni$, i>=$\phi$]  (0,3)
        to [R=${\cal R}_{c}(\phi)$] (2,3)
        to [R] (2,0) to [short] (0,0);
        \draw (1.4,2.1) node{${\cal R}_{pm}$};
        \draw (0.3,3.25) rectangle (1.7,2.7);
        \draw (2,0) to [short] (3,0) 
        to [american current source,invert]  (3,3)
        to [short] (2,3);
        \draw (2.5,2.1) node{$\Phi_{pm}$};
         \draw (4,0) to [V, invert, a=$Ni$, i>=$\phi$]  (4,3)
         to [R=${\cal R}_{c}(\phi)$] (5.5,3)
         to [R=${\cal R}_{pm}$] (7,3) 
         to [V, invert, a=$F_{pm}$]  (7,0)
         to [short] (4,0);
         \draw (4.1,3.25) rectangle (5.4,2.7);
         \draw (0,-4) to [V, invert, a=$Ni$, i>=$\phi$]  (0,-1)
         to [R=${\cal R}_{tot}(\phi)$] (3,-1)
         to [V, invert, a=$F_{pm}$]  (3,-4)
         to [short] (0,-4);
         \draw (0.8,-0.75) rectangle (2.2,-1.25);
        \draw (4,-1) to [short, o-, i=$i$] (5.5,-1)
        to [L, i=$h(\lambda)$] (5.5,-4)
        to [short, -o] (4,-4);
        \draw (5.3,-2) rectangle (5.8,-3);
       
        \draw (5.5, -4) to [short] (7,-4)
        to [american current source, a=$I_{pm}$] (7, -1)
        to [short] (5.5,-1);
        \draw[european voltages] (4.5,-1) to [open, invert, v=$\lambda$] (4.5,-4);

    \end{circuitikz}
    \caption{Derivation of proposed magnetics model for single-coil. Top left: magnetic circuit model. Top right: Thevenin equivalent of permanent magnet. Bottom left: Combination of magnet and core reluctances. Bottom right: Resulting electric circuit model.}
    \label{fig:magckt}
\end{figure}

The motivation for the proposed model is illustrated by considering the single-coil magnetic model  shown in Fig. \ref{fig:magckt}.
\begin{figure}
\begin{center}
  \begin{tikzpicture}[scale=0.75]
        \filldraw[iron!70]
        (0,3) -- (2,3) -- (2,1) -- (5,1) -- (6,2) -- (6,-2) -- (5,-1)
        -- (2,-1) -- (2,-4) -- (0,-4) -- (0,3);
        \draw (2,3) -- (2,1) -- (5,1) -- (6,2) -- (6,-2) -- (5,-1)
        -- (2,-1) -- (2,-3);
        \draw (0,3) -- (0,-3);
        \foreach \x in {2.3, 2.9, 3.5, 4.1, 4.7}
      {
        \draw[fill=yellow] (\x, 1.3) circle (0.2);
        \draw[fill=yellow] (\x, -1.3) circle (0.2);
        }
        \draw (3.5,1.8) node{$N$ Turns};
        \draw (3.5, 0.7) node{Stator Core};
    \draw[fill=violet!70]
    (7,2) -- (9,2) -- (9,-2) -- (7,-2) -- (7,2);
    \draw (7.4,0) node[node font = \Huge]{S};
    \draw (8.6,0) node[node font = \Huge]{N};
    \draw (8,1.7) node{Magnet};
    \filldraw[iron!70]
    (9,3) -- (11,3) -- (11,-4) -- (9,-4) -- (9,3);
    \draw (9,3) -- (9,-3);
    \draw (11,3) -- (11,-3);
    \draw (10,2) node{Rotor};
    \draw (10,1.5) node{Core};
    \draw[dashed]
    (1,0) -- (10,0) -- (10,-3.5) -- (1,-3.5) -- (1,0);
    \draw (5.5,-3.2) node{Conceptual Return Path};
    \draw[-Stealth] (2.5,0) -- (4.5,0);
    \draw (3.5, -0.3) node{$\phi$};

    \end{tikzpicture}
    \caption{Simplified, single-coil magnetics model of PMSM machine}
    \end{center}  
    \label{magdevice}
\end{figure}
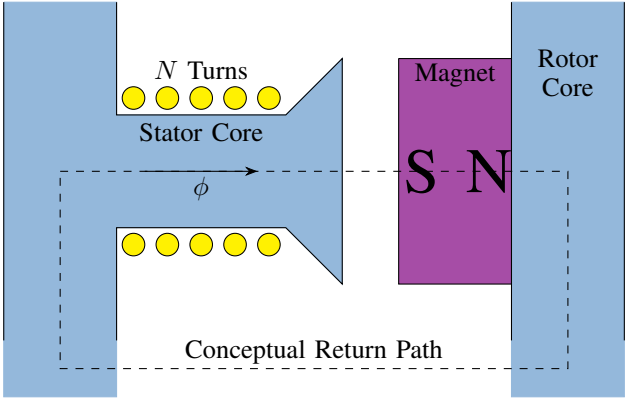
We begin by creating a magnetic circuit model. The $N$-turn coil is represented by an magnetomotive force (MMF) source  $Ni$, where $i$ is the current flowing in the coil. The stator and rotor cores consist of a soft magnetic material exhibiting magnetic saturation, and so are represented by a nonlinear reluctance ${\cal R}_{c}(\phi)$ whose value varies with flux level $\phi$. The PM is assumed to possess linear magnetic properties with an incremental permeability $\mu_m$ and a remanent flux density $B_r$ that varies with rotor temperature $T_r$, and so is modeled as a constant reluctance
${\cal R}_{pm}$ and flux source $\Phi_{pm} (T_r) $, as shown in Fig. \ref{fig:magckt}. 

An equivalent magnetic circuit can be created using the Thevenin equivalent of the PM, where the PM excitation is modelled as MMF source $F_{pm} (T_r) = {\cal R}_{pm} \Phi_{pm} (T_r)$ rather than a flux source. 
Simplifying the magnetic circuit further, the two
reluctances are now in series and can be combined into
a single nonlinear reluctance, ${\cal R}_{tot}(\phi)$. Analysis of the resulting magnetic circuit yields:
\begin{align}
    Ni + F_{pm}(T_{r}) &= {\cal R}_{tot}(\phi)\phi = f(\phi) = f\left(\frac{\lambda}{N} \right);\\
   \Rightarrow i &= \frac{1}{N}\left(f\left(\frac{\lambda}{N} \right) - F_{pm}(T_{r}) \right)\\ 
     &= \textit{h}(\lambda) - I_{pm}(T_{r}).
    \label{prpsdh_eq}
\end{align}
This expression can be represented in an electrical circuit model as a nonlinear inductance, with current/flux-linkage relationship $i=h(\lambda)$, in parallel with a current source $I_{pm} (T_r)$ representing the PM excitation, as shown in Fig.  \ref{fig:magckt}. As the value of $I_{pm}$ is directly proportional to the remanent flux density $B_r$ of the magnet, it will share the same dependence upon the rotor temperature.

In order to illustrate the advantages of the proposed model, we consider a model consisting of a nonlinear inductance and
the more commonly used flux-linkage source representing permanent magnet excitation.
\begin{align}
   \lambda &= {\bf L}(i) i + \Lambda{}_{pm} (T_{r}).
   \label{linmdleq}
\end{align}
The key distinction between the two models is how the current/flux-linkage relationships react to changes in rotor temperature. In the case of the model (\ref{linmdleq}) with a PM flux-linkage source, the result of a rotor temperature change would be a constant offset $\Delta \Lambda$ in the $\lambda-i$ curve. In the case of the proposed model a change in rotor temperature would result in a constant offset $\Delta I$.

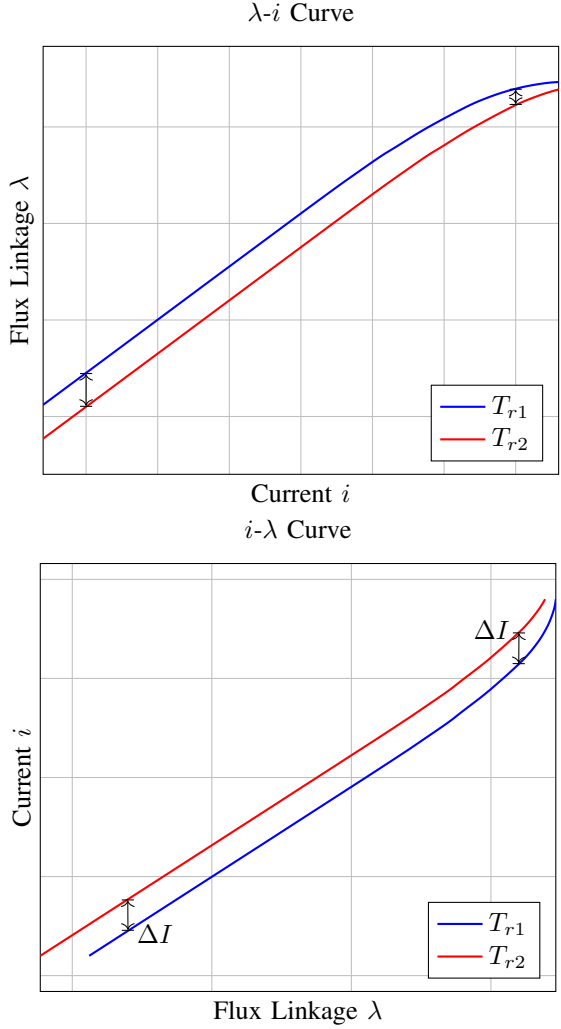
\begin{figure}[htbp]
\begin{center}
\begin{tikzpicture}
\begin{axis}[ticks=none, grid=major, enlarge x limits=false, xscale=1,
xlabel=Current $i$, title=$\lambda$-$i$ Curve,
ylabel=Flux Linkage $\lambda$,
legend entries={$T_{r1}$,$T_{r2}$},
legend style={
legend pos=south east,
},
]
\addplot[smooth, blue, thick] table {li1curve.dat};
\addplot[smooth, red, thick] table {li2curve.dat};
\draw[|<->|] (axis cs: -150,.005001) -- (axis cs: -150,0.0225269);
\draw[|<->|] (axis cs: 150,.16136) --  (axis cs: 150,0.169774);
\end{axis}
\end{tikzpicture}
 \quad
\begin{tikzpicture}
\begin{axis}[ticks=none, grid=major, enlarge x limits=false, xscale=1,
ylabel=Current $i$, title=$i$-$\lambda$ Curve,
xlabel=Flux Linkage $\lambda$,
legend entries={$T_{r1}$,$T_{r2}$},
legend style={
legend pos=south east,
}
]
\addplot[smooth, blue, thick] table {il1curve.dat};
\addplot[smooth, red, thick] table {il2curve.dat};
\draw[|<->|] (axis cs: 0.0199472,-154.775) -- node[below right] {$\Delta I$} (axis cs: 0.0198816,-123.063);
\draw[|<->|] (axis cs: 0.159963,114.414) -- node[above left] {$\Delta I$} (axis cs: 0.160053,146.486);
\end{axis}
\end{tikzpicture}
\end{center}
\caption{Representative flux-linkage/current relationship capturing magnetic saturation at two different rotor temperatures. (a) $\lambda-i$ curve. (b) $i-\lambda$ curve.}
\label{mdlcmp1}
\end{figure}

Figure \ref{mdlcmp1} shows a representative flux linkage/current relationship at two different rotor temperatures. One figure plots flux linkage versus current, the other current versus flux linkage.
As can be seen, the variation of the $\lambda-i$ curve 
with temperature is not captured well by a constant offset in the flux linkage. This is because the saturation effect results in a convergence of the flux linkage values at high current excitation. However, the variation of the $i-\lambda$ curve is well-captured by a constant offset in current.



\subsection{Proposed Permanent Magnet Synchronous Machine Model}
The proposed equivalent two-phase model for a Permanent Magnet Synchronous Machine (PMSM) is developed considering the standard smooth airgap model (i.e., effects due to the slots in the stator core are neglected). 

\subsubsection{Current/flux-linkage relationship}
The current/flux-linkage relationships of the proposed machine model are presented in the rotor reference frame (designated by a superscript 'r'),
which is related to the variables in the stationary reference frame as follows:
\begin{align}
\vec{x}{}^r = e^{-{\bf J}\theta_{re}} \vec{x},  \quad
\vec{x} = e^{{\bf J}\theta_{re}} \vec{x}{}^r.
\end{align}
where $\theta_{re}=\frac{N_{p}}{2} \theta_r$ is the electrical rotor angle, 
\begin{align}
\textbf{J} = \left[ \begin{array}{cc} 0 & -1 \\ 1 & 0 \end{array} \right] \nonumber
\end{align}
is the $90^{\circ}$ rotation matrix, $N_{p}$ is the number of poles of the machine, and $\theta_r$ is the mechanical angle of the rotor. The matrix exponentials $e^{-{\bf J}\theta_{re}}$ and $e^{{\bf J}\theta_{re}}$ represents the Park and Inverse Park Transforms, respectively.

The current/flux-linkage relationships are an extension of the single-coil model discussed in the prequel to a vector representation:
\begin{align}
\label{e7}
\vec{i}{}^r &= \vec{h} \left(\vec{\lambda}{}^r \right) - \vec{i}{}^r_{pm} (T_r)
\end{align}
or
\begin{align}
 \begin{bmatrix}
i^r_{d}\\
i^r_{q}
\end{bmatrix}
&=
 \begin{bmatrix}
h_{d}(\lambda^r_{d},\lambda^r_{q})\\
h_{q}(\lambda^r_{d},\lambda^r_{q})
\end{bmatrix}
-
 \begin{bmatrix}
I_{pm} (T_r) \\
0
\end{bmatrix}
 \end{align}
where $I_{pm}$ represents the permanent magnet excitation.

The mapping $\vec{h} \left(\vec{\lambda}{}^r\right)$ captures the nonlinear effects of saturation of soft magnetic materials of the machine. and  has the following property
\begin{align}
\vec{h}(\vec{0})=\vec{0}.
\end{align}
If the electric machine design has the standard symmetries, $\vec{h} \left(\vec{\lambda}{}^r\right)$ also has the following properties:
\begin{align}
h_q(\lambda^r_{d},-\lambda^r_{q}) = -h_q(\lambda^r_{d},\lambda^r_{q}),\\
h_q(-\lambda^r_{d},\lambda^r_{q}) = h_q(\lambda^r_{d},\lambda^r_{q}),\\
h_d(\lambda^r_{d},-\lambda^r_{q}) = h_d(\lambda^r_{d},\lambda^r_{q}).
\end{align}
Assuming standard magnetic material properties (e.g., incremental permeabilities are always positive), it can be shown that $\vec{h} \left(\vec{\lambda}{}^r\right)$ is injective and surjective, and so an inverse flux-linkage/current relationship exists:
\begin{align}
    \vec{\lambda}{}^r= \vec{h}^{-1} \left(\vec{i}{}^r + \vec{i}{}_{pm}^r\right) = \vec{g}\left(\vec{i}{}^r + \vec{i}{}_{pm}^r \right).
\end{align}

\subsubsection{Stator voltage dynamic equations in rotor reference frame}
It can be shown that:
\begin{align}
\label{e10}
\vec{v}{}^r &= R\vec{i}{}^r + \frac{N_{poles}}{2} \omega_r \textbf{J}\vec{\lambda}^r + \frac{d\vec{{\lambda}}^r}{dt} \\
&= R\left[\vec{h}\left(\vec{\lambda}^r\right) - \vec{i}^r_{pm}\right] + \frac{N_{p}}{2} \omega_r \textbf{J}\vec{\lambda}^r + \frac{d\vec{{\lambda}}^r}{dt},
\end{align}
where $\omega_r$ is the angular velocity of the rotor.

\subsubsection{Electromagnetic Torque}
To derive the torque expression, we begin with an expression for the stored magnetic field energy $w_{fld}$ in a three-phase machine  \cite{fitzgerald}. To simplify the analysis we leave out a (position-independent) energy component due to the magnets, which does not impact the final result.
\begin{align}
w_{fld} \left(\vec{\lambda},\theta_r \right) &= \frac{3}{2}\int_{0}^{\vec{\lambda}} 
 \vec{i}{}^T \left(\vec{x})\right) d\vec{x} \nonumber\\
&= \frac{3}{2}\int_{0}^{\vec{\lambda}}
\left\{  e^{{\bf J} \theta_{re}} \left[ \vec{h} \left( e^{-{\bf J} \theta_{re}}
\vec{x} \right) - \vec{i}{}^r_{pm} \right] \right\}^T d\vec{x}.
\end{align}
The electromagnetic torque can be shown to be:
\begin{align}
\tau_{em} &= -\frac{\partial w_{fld} \left(\vec{\lambda},\theta_r \right) }{\partial \theta_{r}} \nonumber\\
&= \frac{3 N_{p} }{4} \left[ - \left( \vec{\lambda}{}^r \right)^T {\bf J} \vec{h} \left( \vec{\lambda}{}^r \right) +
\vec{\lambda}{}^r {\bf J} \vec{i}{}_{pm}^r \right] \nonumber\\
&= \frac{3 N_{p}}{4} \left[ h_q\left(\vec{\lambda}{}^r \right) \lambda_d^r - h_d\left(\vec{\lambda}{}^r \right) \lambda_q^r +
I_{pm} \left(T_r\right) \lambda_q^r 
\right].
\label{e12}
\end{align}

The first term in (\ref{e12}) corresponds to a (nonlinear) reluctance torque and the second term is the PM torque. 

\section{Validation and Comparison Results}
In this section we will illustrate the accuracy of the proposed model through comparison with a "baseline" PM flux-linkage model. We will first introduce the baseline model. We will then compare how changes in the rotor temperature affect the accuracy of the proposed and baseline models. The first comparisons will be conducted using FEA results for three different machine designs: an interior permanent machine, a distributed-winding surface mount permanent magnet (SMPM) machine, and a concentrated-winding SMPM machine. Finally, a comparison will conducted using experimental results.
\subsection{Baseline model}
We will use the following nonlinear PM flux linkage model from literature \cite{6349974} as a baseline for comparison with the proposed model:
\begin{align}
\label{e3}
    \vec{\lambda}{}^r = {\bf L}(\vec{i}{}^r) \vec{i}{}^r + \vec{\lambda}{}^r_{pm} (T_r),
\end{align}
    \begin{align}
    \label{e4}
    {\bf L}(\vec{i}{}^r) &= \left[ \begin{array}{cc} L_d(i^r_d) & L_{dq}(i^r_d,i^r_q) \\ L_{qd}(i^r_d,i^r_q) & L_q(i^r_q) \end{array} \right], \\ \vec{\lambda}{}^r_{pm} (T_r) &= \left[ \begin{array}{c} \Lambda_{PM} (T_r) \\ 0 \end{array} \right] \nonumber
    \end{align}
where,
  \begin{align}
      L_d(i^r_d) &= \frac{\lambda_d^r(i_d^r,0)-\Lambda_{pm} (T_r)}{i_d^r}, \nonumber\\
      L_q(i^r_q) &= \frac{\lambda_q^r(0,i_q^r)}{i_q^r},\nonumber\\
      L_{dq}(i^r_d,i^r_q) &= \frac{\lambda_d^r(i_d^r,0)-\lambda_d^r(i_d^r,i_q^r)}{i_q^r},\nonumber\\
      L_{qd}(i^r_d,i^r_q) &= \frac{\lambda_q^r(0,i_q^r)-\lambda_q^r(i_d^r,i_q^r)}{i_d^r}\nonumber
 \end{align} 
 and the torque is given by
\begin{gather}
\tau_{em} = \frac{3N_p}{4} \Big[ L_{dq}(i^r_d,i^r_q)\left(i_q^r\right)^2 
 + \left(L_d(i^r_d)-  L_q(i^r_q)\right)i_d^ri_q^r +  \nonumber\\
 \Lambda_{pm}(T_r)i_q^r - L_{qd}(i^r_d,i^r_q)\left(i_d^r\right)^2\Big]. 
\end{gather}
This particular model was chosen as it separates the nonlinear behavior from a rotor-temperature-dependent parameter ($\Lambda_{pm} (T_r)$), the same as the proposed NL PM current model.
    
\subsection{Model comparison using FEA-based data}
\label{SubSection2}
We first use two-dimensional finite element analysis (FEA) results from ANSYS as the basis for comparison. The  process that is used to compare the two models is as follows:
\begin{enumerate}
    \item Using FEA, calculate the flux linkages $\lambda_d^r$ and $\lambda_q^r$ and the electromagnetic torque $\tau_{em}$ over a range of currents $i_d^r$ and $i_q^r$ with the rotor temperature set at $T_1=60^{\circ}C$.
    \item Use the flux-linkage/current data to determine the values of $I_{pm}(T_1)$ (used in the proposed model) and $\Lambda_{pm}(T_1)$ (used in the baseline model). The value of $\Lambda_{pm}$ is determined from $\lambda_d^r$ by setting the currents to zero, and the value of $I_{pm}$ is determined by finding the value of $i_d^r$ that sets $\lambda_d^r=0$.
    \item Generate a look-up table for the $\vec{\lambda}{}^r$=$\vec{g}\left(\vec{i}{}^r+\vec{i}{}_{pm}^r \right)$ mapping of the proposed model using $20^{\circ}C$ FEA data. Note that we determine the $\vec{g}\left(\circ\right)$ mapping rather than the $\vec{h}\left(\circ\right)$ mapping to better enable the comparison. 
    \item Generate a look-up table for the inductances in (\ref{e4}) for the baseline model using the $20^{\circ}C$ FEA data.    
    \item Repeat steps 1) and 2) for a different rotor temperature ($T_2$).
    \item Compute the flux linkages and torques with the proposed model using the nonlinear look-up table for $\vec{g}(\cdot)$ determined from the $T_1$ data and $I_{pm} (T_2)$ for the proposed model.
    \item Compute the flux linkages and torques with the comparison model using the nonlinear look-up table ${\bf L} (\cdot)$ from the $T_1$ data and $\Lambda_{pm} (T_2)$ for the baseline model.
    \item Compare the results from the previous steps 6) and 7) with the FEA results at the new rotor temperature ($T_2$).
\end{enumerate}
\subsubsection{Interior permanent magnet machine}
\begin{figure}[htbp]
    \centering
    \includegraphics[width=.85\linewidth,trim={0cm 0cm 0cm 0cm},clip]{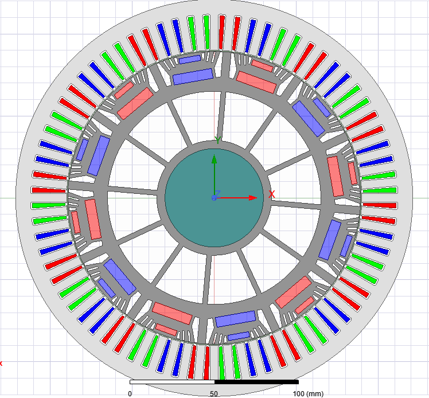}\vspace{-10pt}
    \caption{Cross-section of Interior Permanent Magnet machine \cite{en14092343}}
    \label{ipmmc}
\end{figure}

We first consider the interior permanent magnet machine design presented in \cite{en14092343} and shown in Fig. \ref{ipmmc}.
The permanent magnet material used is NdFeB N35UH from Arnold Magnetics \cite{N35UH}. The mechanical dimensions and winding parameters are kept the same as in \cite{en14092343}. With the recent advancements in cooling mechanisms as shared in \cite{9923908}, peak current densities of 48 A$_{rms}$/mm$^2$ can be achieved. We use this value to determine the maximum current of the mapping, though we note that this machine was not specifically designed for these current densities.

\begin{figure}[htbp]
    \centering
    \includegraphics[width=1\linewidth,trim={0cm 0cm 0cm 0cm},clip]{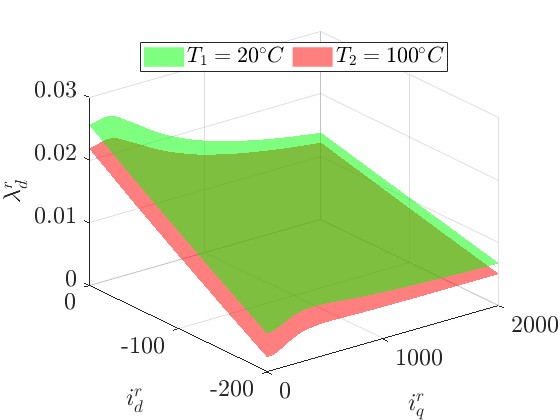}
    \quad
    \includegraphics[width=1\linewidth,trim={0cm 0cm 0cm 0cm},clip]{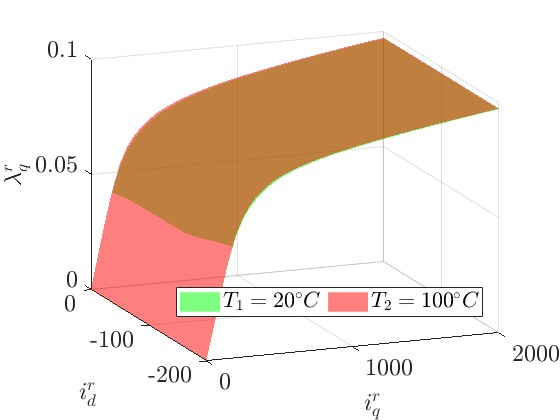}
    \caption{$\vec{\lambda}^r$ - $\vec{i}^r$ mapping of interior permanent magnet machine at different rotor temperatures}
    \label{ipmgfunc}
\end{figure}

The NL $\vec{\lambda}^r$ - $\vec{i}^r$ mapping, determined using FEA at temperatures $T_1$=$20^{\circ}C$ and $T_2$=$100^{\circ}C$, is shown in Fig. \ref{ipmgfunc}. 
As is to be expected, we see a significant change in $\lambda_d^r$ whereas $\lambda_q^r$ is not substantially different.
The value of $I_{pm}$ and $\Lambda_{pm}$ was determined for multiple temperatures and is shown in Table. \ref{Ipmlamvalipm}.
\begin{table}[htbp]
\begin{center}
\caption{$I_{pm}$ and $\Lambda_{pm}$ values for Interior Permanent Magnet machine}
\label{Ipmlamvalipm}
\begin{tabular}{ |c|c|c| } 
 \hline
 Temperature ($^{\circ}C$) & $I_{pm}$ (A) & $\Lambda_{pm} (V-s)$\\
 \hline
20 & 264.8 & 0.0256\\ 
60 & 245.6 & 0.0238\\ 
80 & 235.1 & 0.0227\\ 
100 & 225.4 & 0.0219\\ 
   \hline
\end{tabular}
\end{center}
\end{table}

\begin{figure}[htbp]
    \centering
    \includegraphics[width=1.15\linewidth,trim={0cm 0cm 0cm 0cm},clip]{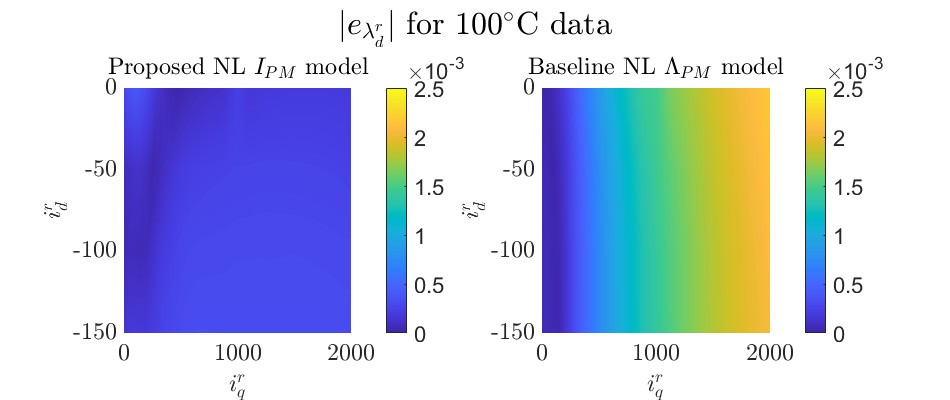}\vspace{-10pt}
    \caption{Absolute errors in $\lambda_d^r$  for proposed and baseline models of Interior Permanent Magnet machine when compared to FEA results.}
    \label{absldripm}
\end{figure}

\begin{figure}[htbp]
    \centering
    \includegraphics[width=1.15\linewidth,trim={0cm 0cm 0cm 0cm},clip]{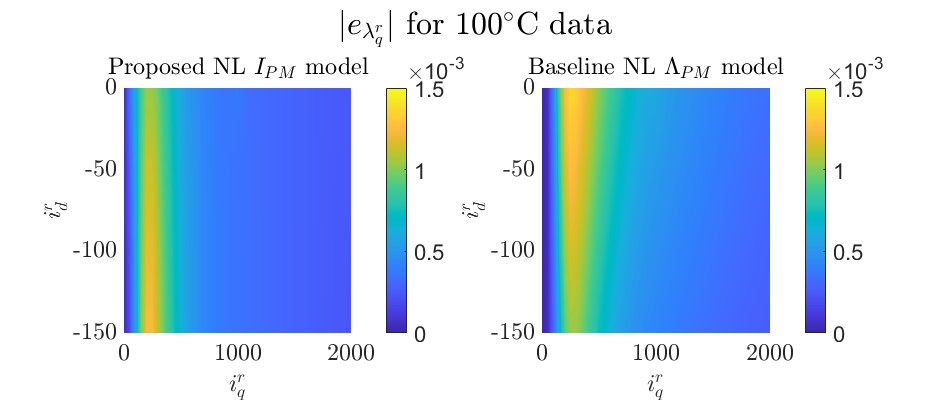}\vspace{-10pt}
    \caption{Absolute errors in $\lambda_q^r$ for proposed and baseline models of Interior Permanent Magnet machine when compared to FEA results.}
    \label{abslqripm}
\end{figure}

  The absolute errors in $\lambda_d^r$ and $\lambda_q^r$ for both the proposed PM current model and the baseline PM flux linkage model are shown in Figs. \ref{absldripm} - \ref{abslqripm}. 
  It can be seen that the baseline PM flux linkage model experiences significantly larger error in $\lambda_d^r$ at high values of $i_q^r$ when compared to the proposed PM current model. In the case of $\lambda_q^r$, the model errors are similar, and occur at the onset of magnetic saturation on the quadrature axis.
  
\begin{figure}[htbp]
    \centering
    \includegraphics[width=1.15\linewidth,trim={0cm 0cm 0cm 0cm},clip]{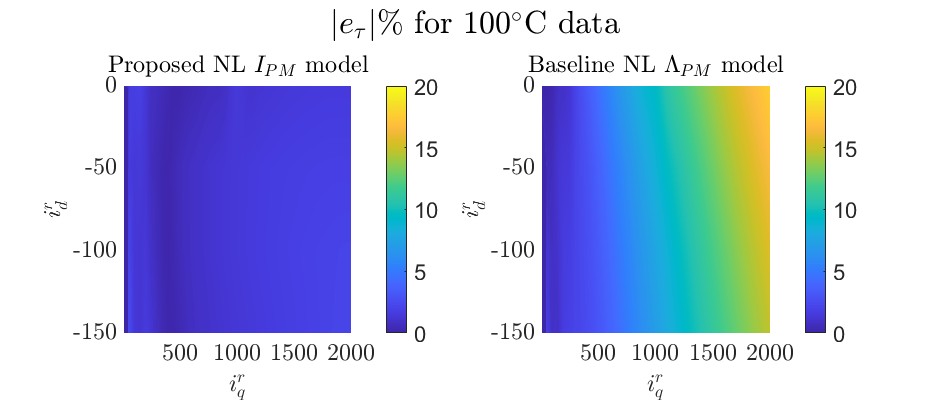}\vspace{-10pt}
    \caption{Relative errors in torque for proposed and baseline models of Interior Permanent Magnet machine when compared to FEA results.}
    \label{reltrqipm}
\end{figure}

  The relative error in torque for the two models is highlighted in Fig. \ref{reltrqipm}. 
It can be clearly seen that the proposed PM current model shows a significant improvement in torque accuracy when compared to the PM flux linkage model, especially at high values of $i_q^r$ when the  machine iron is heavily saturated.

\subsubsection{Distributed-winding SMPM machine}
\begin{figure}[htbp]
    \centering
    \includegraphics[width=.85\linewidth,trim={0cm 0cm 0cm 0cm},clip]{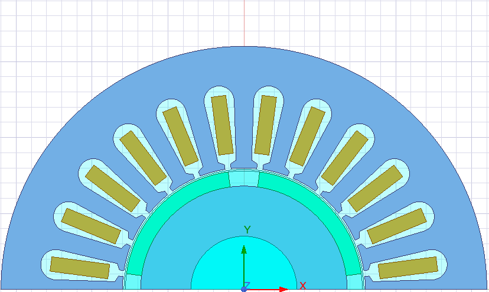}\vspace{-10pt}
    \caption{Cross-section of Distributed winding SMPM machine in ANSYS Maxwell}
    \label{dismc}
\end{figure}

\begin{figure}[htbp]
    \centering
    \includegraphics[width=1.025\linewidth,trim={0cm 0cm 0cm 0cm},clip]{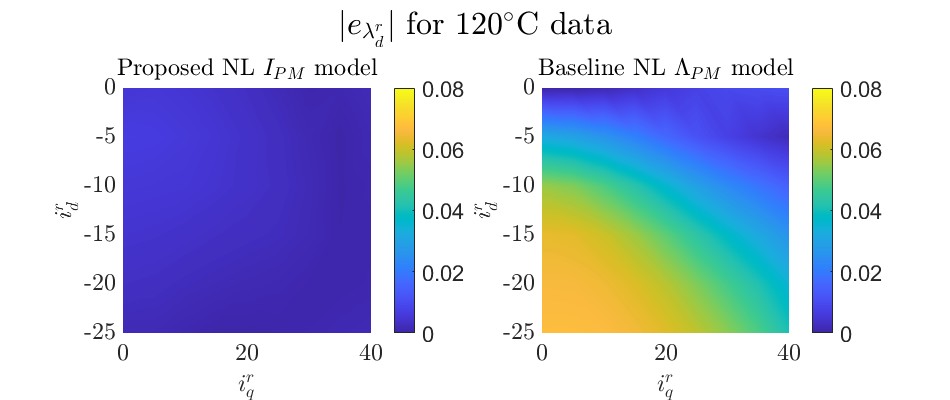}\vspace{-5pt}
    \caption{Absolute errors in $\lambda_d^r$  for proposed and baseline models of distributed winding SMPM}
    \label{absldr}
\end{figure}

\begin{figure}[htbp]
    \centering
    \includegraphics[width=1.05\linewidth,trim={0cm 0cm 0cm 0cm},clip]{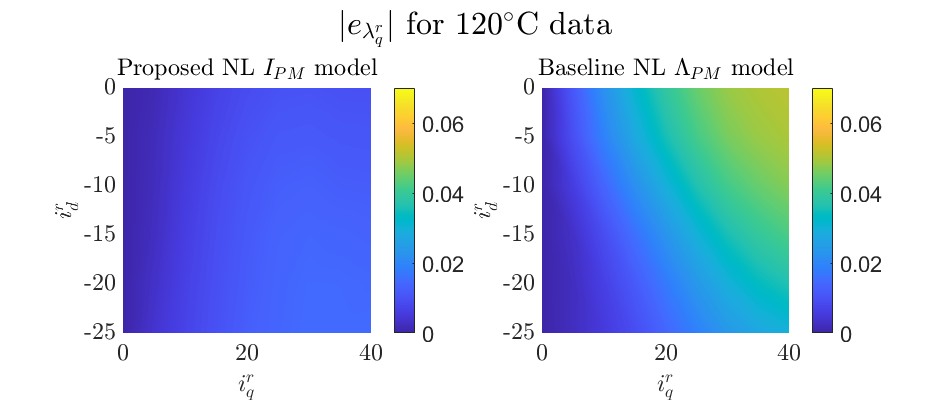}\vspace{-10pt}
    \caption{Absolute errors in $\lambda_q^r$  for proposed and baseline models of distributed winding SMPM}
    \label{abslqr}
\end{figure}

\begin{figure}[htbp]
    \centering
    \includegraphics[width=1.05\linewidth,trim={0cm 0cm 0cm 0cm},clip]{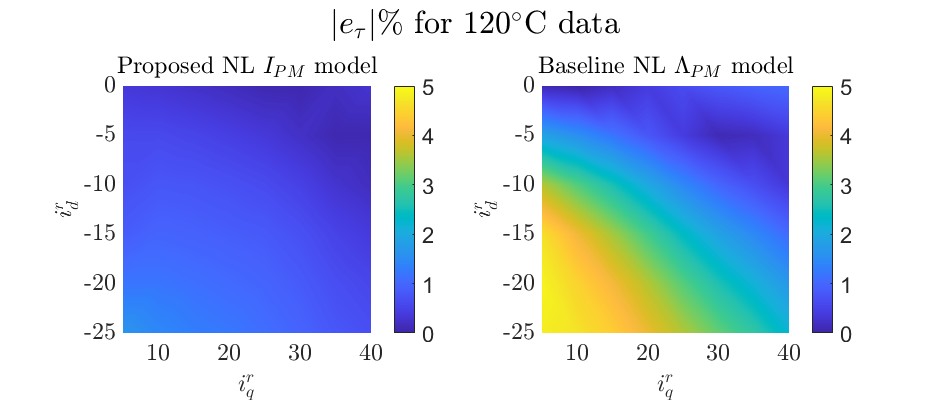}\vspace{-10pt}
    \caption{Relative errors in torque for proposed and baseline models of distributed winding SMPM}
    \label{reltrq}
\end{figure}

A 4 pole, distributed winding SMPM (Fig. \ref{dismc}) with machine parameters as shown in Table. \ref{dismcparams} is simulated in ANSYS FEA for rotor temperatures $T_1 = 20^{\circ}C$ and $T_2 = 120^{\circ}C$. The machine is simulated with current density of 13 A$_{rms}$/mm$^2$, which corresponds to forced-air cooling conditions. It can be seen that the proposed PM current model outperforms the PM flux linkage model.
\begin{table}[!h]
\begin{center}
\caption{Distributed winding SMPM parameters}
\label{dismcparams}
\begin{tabular}{ |c|c| } 
 \hline
 Parameter & Value \\ 
 Number of poles & 4 \\ 
 Number of slots & 24 \\ 
 Conductors per slot & 100\\
 Parallel branches & 1\\
 Magnet material & NdFe35 (N35UH)\\
 Steel type & M19 24G\\
 Stator outer diameter (mm) & 160\\
 Stator inner diameter (mm) & 80\\
 Rotor outer diameter (mm) & 78\\
 Magnet thickness (mm) & 5\\
 Machine length (mm) & 80\\
 \hline
\end{tabular}
 \label{dismcparams}
\end{center}
\end{table}
Values of $I_{pm}$ and $\Lambda_{pm}$ for different rotor temperatures are provided in Table. \ref{Ipmlamvaldis} for this machine design. 
\begin{table}[htbp]
\begin{center}
\caption{$I_{pm}$ and $\Lambda_{pm}$ values for Distributed winding SMPM machine}
\label{Ipmlamvaldis}
\begin{tabular}{ |c|c|c| } 
 \hline
 Temperature ($^{\circ}C$) & $I_{pm}$ (A) & $\Lambda_{pm} (V-s)$\\
 \hline
20 & 40.38 & 1.2643\\ 
60 & 38.4 & 1.2281\\ 
80 & 37.3 & 1.2048\\ 
100 & 36.3 & 1.1842\\
120 & 35.1 & 1.156\\
   \hline
\end{tabular}
\end{center}
\end{table}
Figs \ref{absldr} through \ref{reltrq} show absolute errors in $\lambda_d^r$ and $\lambda_q^r$ and the relative torque error for the two models at temperature $T_2$. It can be seen that, for $\lambda_d^r$ and hence torque, significant errors occur for negative values of $i_d^r$. This is due to the fact that the machine iron is saturated by the permanent magnets at zero current excitation. The field-weakening effect of negative $i_d^r$ drives the iron out of saturation and so causes the effective PM flux linkage to increase, resulting in the error.
\subsubsection{Concentrated-winding SMPM machine}
\begin{figure}[htbp]
    \centering
    \includegraphics[width=.85\linewidth,trim={0cm 0cm 0cm 0cm},clip]{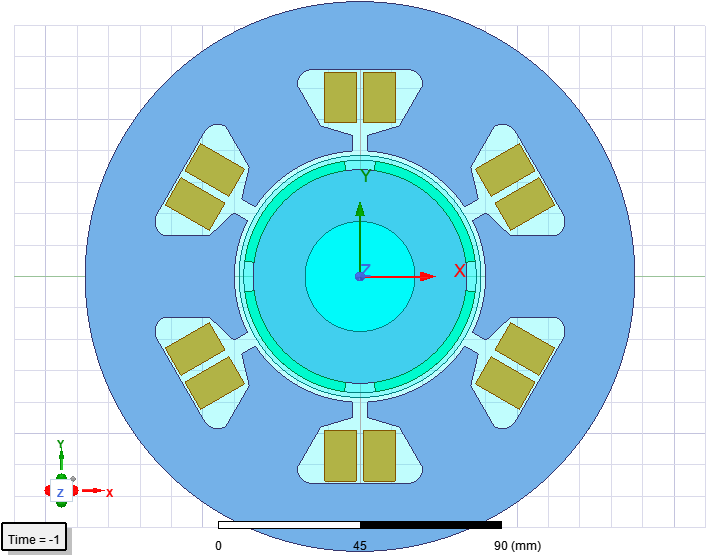}\vspace{-10pt}
    \caption{Cross-section of concentrated winding SMPM machine in ANSYS Maxwell}
    \label{cncmc}
\end{figure}

\begin{figure}[htbp]
    \centering
    \includegraphics[width=1.1\linewidth,trim={0cm 0cm 0cm 0cm},clip]{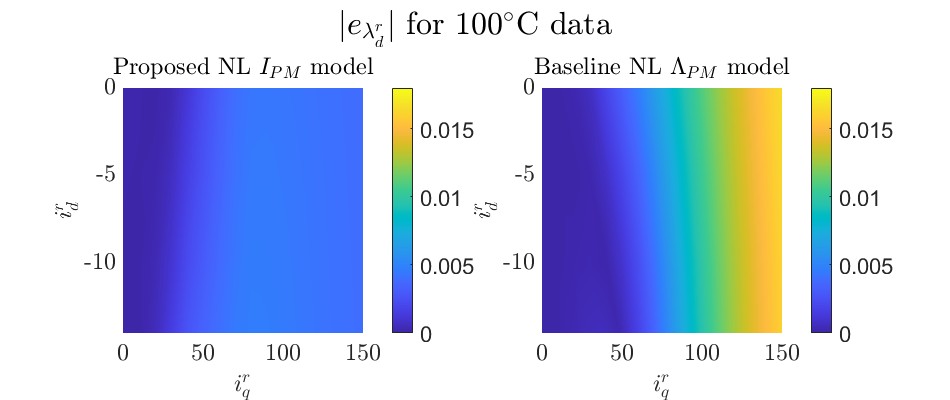}\vspace{-10pt}
    \caption{Absolute errors in $\lambda_d^r$  for proposed and baseline models of concentrated winding SMPM}
    \label{absldidn}
\end{figure}

\begin{figure}[htbp]
    \centering
    \includegraphics[width=1.1\linewidth,trim={0cm 0cm 0cm 0cm},clip]{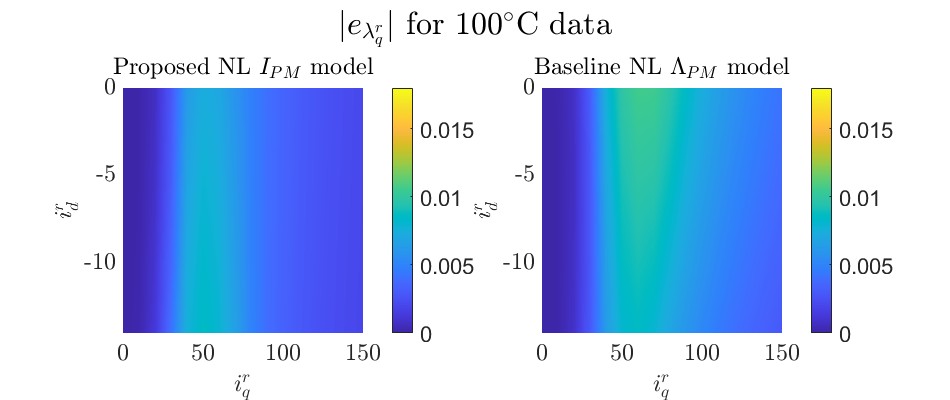}\vspace{-10pt}
    \caption{Absolute errors in $\lambda_q^r$  for proposed and baseline models of concentrated winding SMPM}
    \label{abslqidn}
\end{figure}

\begin{figure}[htbp]
    \centering
    \includegraphics[width=1.1\linewidth,trim={0cm 0cm 0cm 0cm},clip]{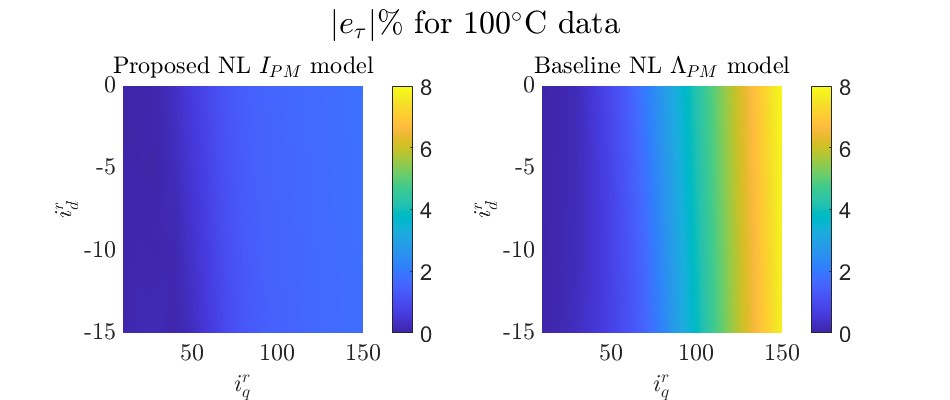}\vspace{-10pt}
    \caption{Relative errors in torque for proposed and baseline models of concentrated winding SMPM}
    \label{reltrqidn}
\end{figure}

A simple 4-pole concentrated winding SMPM with machine parameters as shown in Table. \ref{cncmcparams} is simulated in ANSYS FEA for rotor temperatures $T_1=20^{\circ}C$ and $T_2=100^{\circ}C$. The machine is simulated with a peak current density of 35.5 A$_{rms}$/mm$^2$, corresponding to liquid cooling conditions. The machine cross section is shown in Fig. \ref{cncmc}. 
\begin{table}[!h]
\begin{center}
\caption{Concentrated winding SMPM parameters}
\label{cncmcparams}
\begin{tabular}{ |c|c| } 
 \hline
 Parameter & Value \\ 
 Number of poles & 4 \\ 
 Number of slots & 6 \\ 
 Conductors per slot & 220\\
 Parallel branches & 1\\
 Magnet material & NdFe35 (N35UH)\\
 Steel type & M19 24G\\
 Stator outer diameter (mm) & 220\\
 Stator inner diameter (mm) & 80\\
 Rotor outer diameter (mm) & 74\\
 Magnet thickness (mm) & 3\\
 Machine length (mm) & 80\\
 \hline
\end{tabular}
 \label{cncmcparams}
\end{center}
\end{table}
The $I_{pm}$ and $\Lambda_{pm}$ values for this machine is shown in Table. \ref{Ipmlamvalcnc}.
\begin{table}[htbp]
\begin{center}
\caption{$I_{pm}$ and $\Lambda_{pm}$ values for concentrated winding SMPM machine}
\label{Ipmlamvalcnc}
\begin{tabular}{ |c|c|c| } 
 \hline
 Temperature ($^{\circ}C$) & $I_{pm}$ (A) & $\Lambda_{pm} (V-s)$\\
 \hline
20 & 26.6167 & 0.399\\
60 & 25.3747 & 0.3809\\ 
80 & 24.6345 & 0.3698\\ 
100 & 24.0188 & 0.3605\\
   \hline
\end{tabular}
\end{center}
\end{table}
    \label{ipmtmp}
The absolute error in $\lambda_d^r$ and $\lambda_q^r$ for both models is compared as shown in Figs. \ref{absldidn}-\ref{abslqidn}. Relative error in estimated torque by both models is compared in Fig. \ref{reltrqidn}. We can clearly observe the improvement in accuracy in the proposed model from Figs \ref{absldidn} and \ref{reltrqidn}, especially in the region of high torque where magnetic saturation is most prominent.
\subsection{Experimental Validation}
\label{SubSection4}
\subsubsection{Setup}
\begin{figure}[htbp]
\begin{center}
 \includegraphics[width=.85\linewidth,trim={0cm 0cm 0cm 0cm},clip]{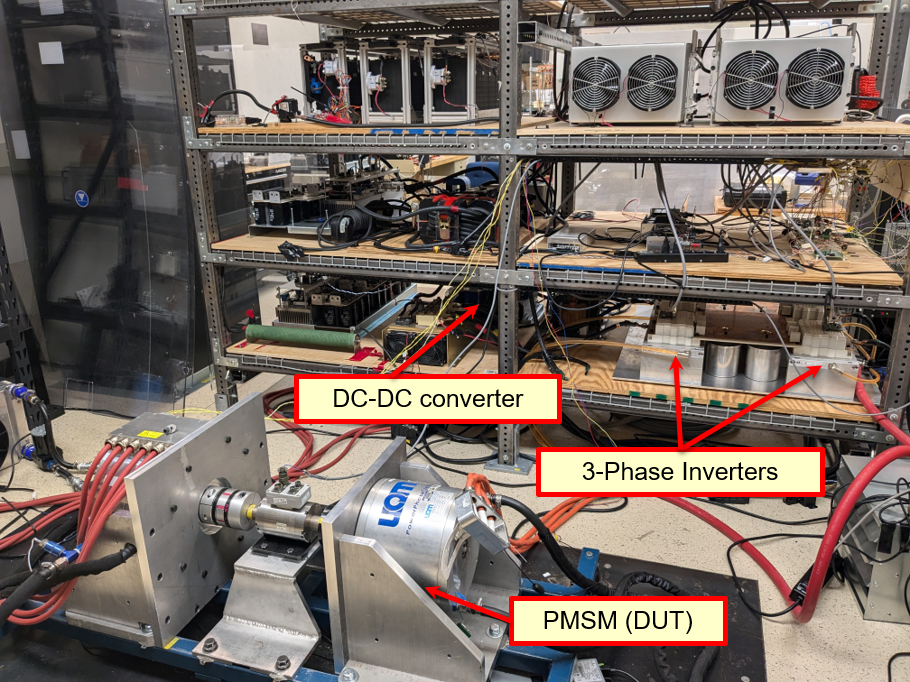}\vspace{-10pt}
 \quad
\includegraphics[width=.85\linewidth,trim={0cm 0cm 0cm 0cm},clip]{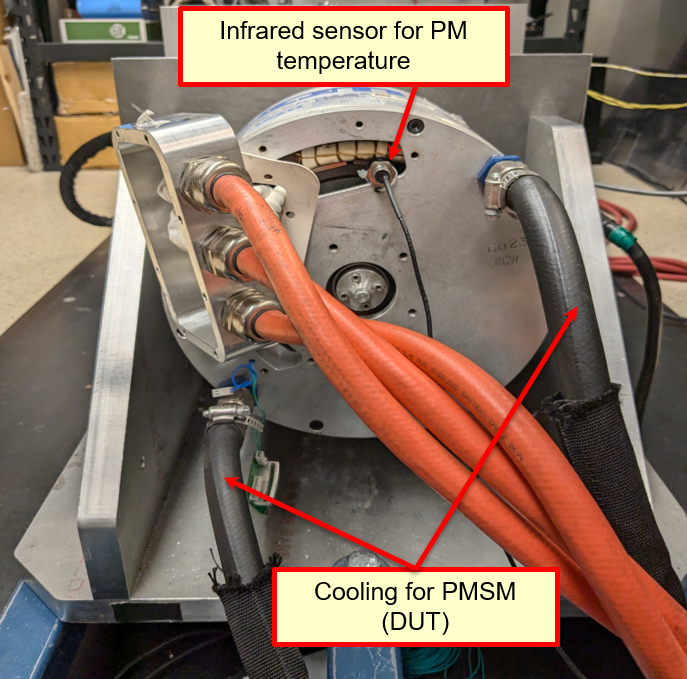}\vspace{-10pt}
\end{center}
\caption{(a) Hardware setup for experiment, (b) Infrared sensor for rotor temperature measurement on the SMPM test machine}
\label{hrdsetupf}
\end{figure}

A three phase, 18 pole, 145kW UQM SMPM with specifications as shown in Table \ref{hrdparams} is characterized using a dSpace Scalexio real-time target controller that uses auto-generated code from MATLAB/Simulink. A 3 phase voltage source inverter consisting of SKiiP 1814 GB17E4-3DUW V2 IGBTs, made by Semikron, is used with a switching frequency of 10 kHz, bus voltage of 175 V, and dead-time of 2 µs. Space-Vector Modulation (SVM) is used as the modulation scheme to calculate the desired duty-cycles. The rotor speed is kept constant at 500rpm by an electric-machine-based dynamometer. This speed was chosen as the voltage/current relationships of the machine are relatively insensitive to both winding resistance and core losses at the corresponding electrical frequency (75Hz).
An infrared temperature sensor from Texense (IRN2) is used to measure and maintain the rotor temperature to constant. Since it is difficult to keep the rotor temperature constant, efforts were made to collect data quickly to keep the temperature variation within 5$^{\circ}$C.  The experimental setup of the hardware is shown in Fig. \ref{hrdsetupf}.
\begin{table}[!h]
\begin{center}
\caption{Test machine rating}
\label{hrdparams}
\begin{tabular}{ |c|c| } 
 \hline
 Description & Value \\ 
 Number of poles & 18 \\ 
 Manufacturer & UQM Technologies \\ 
 Type & PM Brushless\\
 Maximum speed & 8000rpm\\
 Maximum Current & 500A\\
 Maximum Voltage & 450V\\
 Rated Power & 145KW peak/ 85 KW continuous\\
  \hline
\end{tabular}
 \label{dismcparams}
\end{center}
\end{table}

\begin{figure}[htbp]
    \centering
    \includegraphics[width=1\linewidth,trim={0cm 0cm 0cm 0cm},clip]{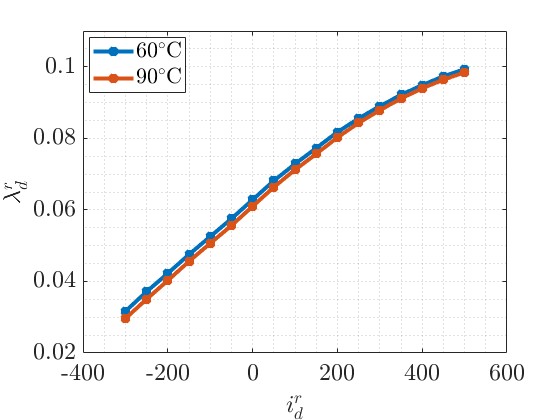}\vspace{-10pt}
    \caption{Experimental $\lambda_d^r$ - $i_d^r$ characterization of UQM 145 machine for rotor temperatures $T_1 = 60^{\circ}C$ and
$T_2=90^{\circ}C$, Test 1.}
    \label{idronly}
\end{figure}

The test SMPM flux-linkage/current relationship was characterized using a process similar to that provided in \cite{Armando2013ExperimentalIO,6939721}, where flux linkage is estimated by numerically integrating the electromotive force (emf) of the stator windings. 
The machine under test does not experience appreciable saturation under normal operation. Therefore, to illustrate the benefits of the proposed model, operating points that bring the machine iron into magnetic saturation were used:
\begin{itemize}
    \item \noindent \underline{\it{Test 1}}:\\
    Current $i_q^r$ was set to zero and  $i_d^r$ was varied from -300A to 500A.
   \item \noindent \underline{\it{Test 2}}:\\
    Current $i_q^r$ was varied from -400A to 400A and  $i_d^r$ was varied from 100A to 300A.
\end{itemize}
The resulting $\lambda_d^r - i_d^r$ relationships for Test 1 are shown in Fig. \ref{idronly} for rotor temperatures $T_1 = 60^{\circ}C$ and $T_2=90^{\circ}C$.

\subsubsection{Calculation of $I_{pm}$ and $\Lambda_{pm}$}
The $\lambda_d^r-i_d^r$ characterization process used does not provide accurate flux-linkage estimates with high levels of field weakening (i.e., negative values of $i_d^r$.) This is due to the fact that, in these circumstances, the emf of the windings becomes a small fraction of the overall voltage. As a result, the estimated flux linkages become highly sensitive to errors in the estimated winding resistance and voltage drops of the inverter transistors, which are used to calculate the emf. Hence we cannot accurately determine the zero-flux-linkage condition as we did with the FEA results, and so another method must be used to determine the value of $I_{pm}$. This is achieved through the following process:
\begin{itemize}
    \item Using the experimentally determined $\lambda_d^r-i_d^r$ data at temperature $T_1$, create a function which calculates $
    \lambda_d^r  =g_{dT_1} \left(i_d^r  + I_{pmT_1} \right)$ for a given value of $I_{pmT_1}$ through the use of lookup tables and linear interpolation.
    \item Repeat the above step for data taken at temperature $T_2$.
    \item Determine the values of $I_{pm}$ at temperatures $T_1$ and $T_2$ simultaneously by solving the following minimization problem
\end{itemize}
   \begin{gather}
 \label{optieq}
\min_{I_{pmT1},I_{pmT2}} \left\Vert g_{dT_1} \left(i_d^r+I_{pmT_1} \right) - g_{dT_2}\left(i_d^r +I_{pmT_2} \right) \right\Vert^2, 
\end{gather}
where $\left\Vert \cdot \right\Vert$ represents the $L^2$ norm.

For the baseline NL PM flux linkage model, $\Lambda_{pm}$ is computed from the experimental through two different methods as described below. \\
\noindent \underline{\it{Method 1}}:
This is same method described earlier in \ref{SubSection2} where $\Lambda_{pm}$ is determined from the estimated $\lambda_d^r$ when the currents to zero.

\noindent \underline{\it{Method 2}}: 
This method is analogous to the one used to compute $I_{pm}$. Functions
$f_{dT_1} \left( \lambda_d^r - \Lambda_{pmT_1} \right)$ and $f_{dT_2} \left( \lambda_d^r - \Lambda_{pmT_2} \right)$ are constructed for given values of $\Lambda_{pmT_1}$ and $\Lambda_{pmT_2}$ using the experimental data taken at temperatures $T_1$ and $T_2$, respectively. The values of the PM flux linkage at temperatures $T_1$ and $T_2$ are then determined simultaneously by solving the following minimization problem
\begin{gather}
 \label{optieqlpm}
 \min_{\Lambda_{pmT_1},\Lambda_{pmT_2}} \left\Vert f_{dT_1} \left( \lambda_d^r - \Lambda_{pmT_1} \right) - f_{dT_2} \left( \lambda_d^r - \Lambda_{pmT_2} \right)  \right\Vert^2.
\end{gather}
The solver fmincon in MATLAB was used to solve the optimization problems to compute $I_{pm}$ and $\Lambda_{pm}$. The values of $I_{pm}$ and $\Lambda_{pm}$ for both rotor temperartures ($T_1 = 60^{\circ}C$ and $T_2=90^{\circ}C$) from experimental results is shown in Table. \ref{Ipmlamval}.
\begin{table}[htbp]
\begin{center}
\caption{Experimental $I_{pm}$ and $\Lambda_{pm}$ values}
\label{Ipmlamval}
\begin{tabular}{ |c|c|c|c| } 
 \hline
 \multirow{2}{*}{Temperature ($^{\circ}C$)} & \multirow{2}{*}{$I_{pm}$ (A)} & \multicolumn{2}{c|}{$\Lambda_{pm} (V-s)$} \\\cline{3-4}
      &   & Method 1 & Method 2\\
 \hline
60 & 603.6 & 0.0628 & 0.0625\\ 
90 & 585.4 & 0.0610 & 0.0609\\ 
   \hline
\end{tabular}
\end{center}
\end{table}
\subsubsection{Comparison results}
\begin{figure}[htbp]
    \centering
    \includegraphics[width=1\linewidth,trim={0cm 0cm 0cm 0cm},clip]{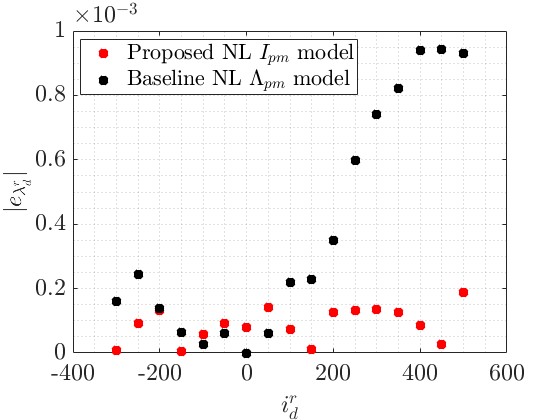}\vspace{-10pt}
    \caption{Absolute errors in $\lambda_d^r$  for proposed NL PM current and baseline NL PM flux linkage models at 90$^{\circ}$C using Method 1 for $\Lambda_{pm}$ calculation for Test 1}
    \label{9060}
\end{figure}

\begin{figure}[htbp]
    \centering
    \includegraphics[width=1\linewidth,trim={0cm 0cm 0cm 0cm},clip]{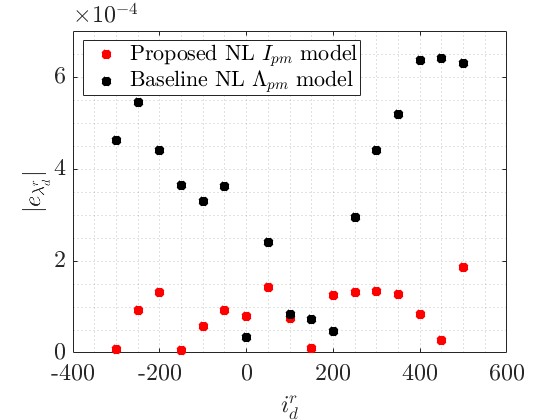}\vspace{-10pt}
   \caption{Absolute errors in $\lambda_d^r$  for proposed NL PM current and baseline NL PM flux linkage models at 90$^{\circ}$C using Method 2 for $\Lambda_{pm}$ calculation for Test 1}
    \label{9080}
\end{figure}

\begin{figure}[htbp]
    \centering
    \includegraphics[width=1.05\linewidth,trim={0cm 0cm 0cm 0cm},clip]{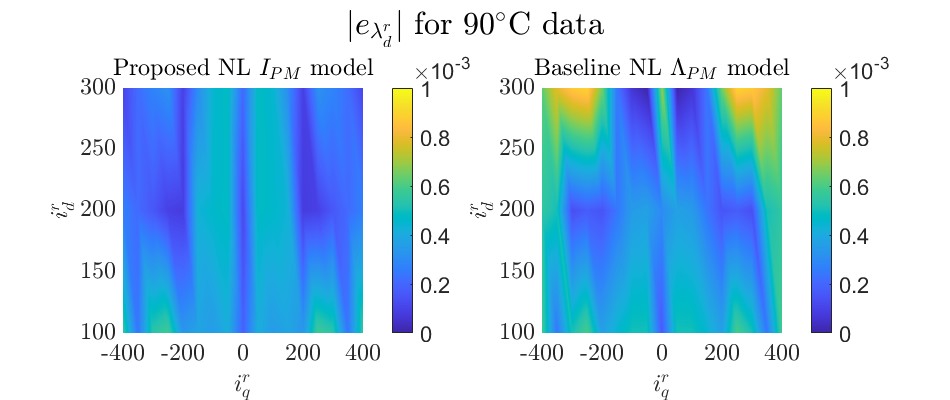}\vspace{-10pt}
   \caption{Absolute errors in $\lambda_d^r$  for proposed NL PM current and baseline NL PM flux linkage models at 90$^{\circ}$C using Method 1 for $\Lambda_{pm}$ calculation for Test 2}
    \label{9081}
\end{figure}

\begin{figure}[htbp]
    \centering
    \includegraphics[width=1.05\linewidth,trim={0cm 0cm 0cm 0cm},clip]{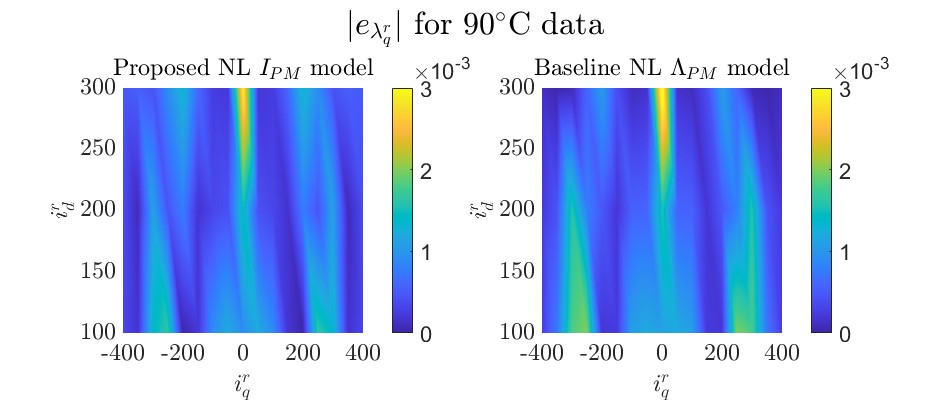}\vspace{-10pt}
   \caption{Absolute errors in $\lambda_q^r$  for proposed NL PM current and baseline NL PM flux linkage models at 90$^{\circ}$C using Method 1 for $\Lambda_{pm}$ calculation for Test 2}
    \label{9082}
\end{figure}

\begin{figure}[htbp]
    \centering
    \includegraphics[width=1.05\linewidth,trim={0cm 0cm 0cm 0cm},clip]{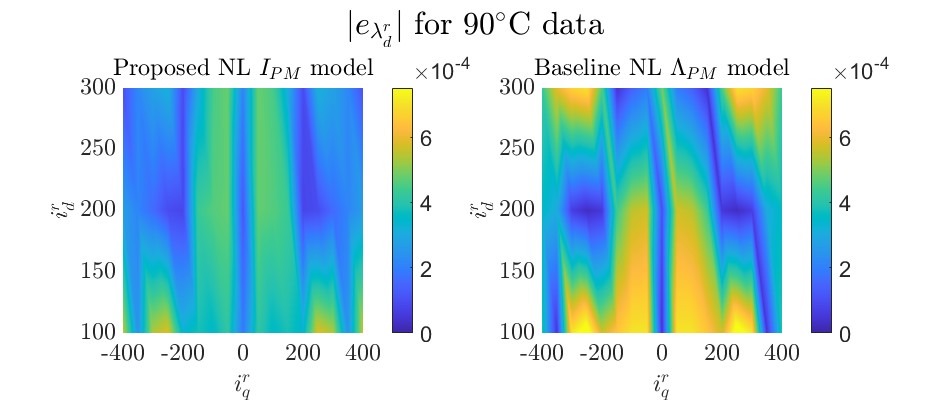}\vspace{-10pt}
   \caption{Absolute errors in $\lambda_d^r$  for proposed NL PM current and baseline NL PM flux linkage models at 90$^{\circ}$C using Method 2 for $\Lambda_{pm}$ calculation for Test 2}
    \label{9083}
\end{figure}

\begin{figure}[htbp]
    \centering
    \includegraphics[width=1.05\linewidth,trim={0cm 0cm 0cm 0cm},clip]{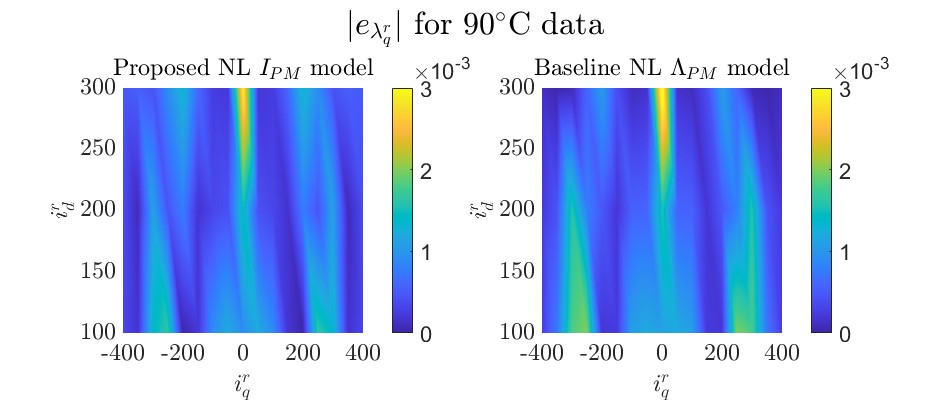}\vspace{-10pt}
   \caption{Absolute errors in $\lambda_q^r$  for proposed NL PM current and baseline NL PM flux linkage models at 90$^{\circ}$C using Method 2 for $\Lambda_{pm}$ calculation for Test 2}
    \label{9084}
\end{figure}

The models were compared using the same technique used in \ref{SubSection2} for the FEA results. Fig. \ref{9060} shows the absolute error in $\lambda_d^r$ for the NL PM current model and NL PM flux linkage model for Test 1, where the value of $\Lambda_{pm}$ for the NL PM flux linkage model at temperature $T_2$ was computed using Method 1. While the models show similar accuracy in the linear region (i.e., negative $i_d^r$ excitation), one can distinctly observe the improved accuracy of the  proposed model in the region of magnetic saturation. 
Fig. \ref{9080} compares the absolute error in $\lambda_d^r$ when $\Lambda_{pm}$ for the NL PM flux linkage model is computed using Method 2 for the same Test 1 condition. In this case, the proposed model shows better accuracy in both the nonlinear and linear regions.
Figs. \ref{9081}-\ref{9084} show the results for Test 2. Fig. \ref{9081} and Fig. \ref{9083} highlight the absolute error in $\lambda_d^r$ for the NL PM current model and NL PM flux linkage model using Method 1 and 2 for $\Lambda_{pm}$ calculation respectively. The proposed model demonstrates higher accuracy in both cases. We observe similar model error in case of $\lambda_q^r$ (Fig. \ref{9082} and Fig. \ref{9084}), which is consistent with the observation from the simulation results.

\section{Conclusion}
\label{Section5}
A machine model for PMSMs that captures the effects of magnetic saturation and rotor temperature variations has been proposed that is accurate and computationally inexpensive, and so can be utilized in real time control applications. The main contribution of the proposed model is the decoupling of the effects of magnetic saturation and rotor temperature variation, the latter being captured in a PM current parameter $I_{pm}$. This parameter can be adaptively estimated, thereby avoiding the need for rotor temperature measurement.  The proposed model is validated using FEA simulation for an interior permanent magnet machine design,  a distributed winding SMPM design, and a concentrated winding SMPM design, and shows high accuracy at different rotor temperatures when compared with a nonlinear PM flux linkage model. Experimental results also show that the proposed model is more accurate than the nonlinear PM flux linkage model. 

Future work will involve the development of a torque regulator algorithm, including an adaptive estimator for PM current, that is based upon the proposed model.

\section{Acknowledgements}
This material is based on work supported by the Office of Naval Research under Grant No.: N00014-18-1-2330.

\bibliographystyle{IEEEtran}
\bibliography{bibligraphy}

\begin{thebibliography}{10}
\providecommand{\url}[1]{#1}
\csname url@samestyle\endcsname
\providecommand{\newblock}{\relax}
\providecommand{\bibinfo}[2]{#2}
\providecommand{\BIBentrySTDinterwordspacing}{\spaceskip=0pt\relax}
\providecommand{\BIBentryALTinterwordstretchfactor}{4}
\providecommand{\BIBentryALTinterwordspacing}{\spaceskip=\fontdimen2\font plus
\BIBentryALTinterwordstretchfactor\fontdimen3\font minus \fontdimen4\font\relax}
\providecommand{\BIBforeignlanguage}[2]{{%
\expandafter\ifx\csname l@#1\endcsname\relax
\typeout{** WARNING: IEEEtran.bst: No hyphenation pattern has been}%
\typeout{** loaded for the language `#1'. Using the pattern for}%
\typeout{** the default language instead.}%
\else
\language=\csname l@#1\endcsname
\fi
#2}}
\providecommand{\BIBdecl}{\relax}
\BIBdecl

\bibitem{4168013}
C.~C. Chan, ``The state of the art of electric, hybrid, and fuel cell vehicles,'' \emph{Proceedings of the IEEE}, vol.~95, no.~4, pp. 704--718, 2007.

\bibitem{7210190}
Z.~Yang, F.~Shang, I.~P. Brown, and M.~Krishnamurthy, ``Comparative study of interior permanent magnet, induction, and switched reluctance motor drives for ev and hev applications,'' \emph{IEEE Transactions on Transportation Electrification}, vol.~1, no.~3, pp. 245--254, 2015.

\bibitem{5764534}
G.~Pellegrino, A.~Vagati, P.~Guglielmi, and B.~Boazzo, ``Performance comparison between surface-mounted and interior pm motor drives for electric vehicle application,'' \emph{IEEE Transactions on Industrial Electronics}, vol.~59, no.~2, pp. 803--811, 2012.

\bibitem{7112507}
B.~Bilgin, P.~Magne, P.~Malysz, Y.~Yang, V.~Pantelic, M.~Preindl, A.~Korobkine, W.~Jiang, M.~Lawford, and A.~Emadi, ``Making the case for electrified transportation,'' \emph{IEEE Transactions on Transportation Electrification}, vol.~1, no.~1, pp. 4--17, 2015.

\bibitem{1211236}
J.~Weimer, ``The role of electric machines and drives in the more electric aircraft,'' in \emph{IEEE International Electric Machines and Drives Conference, 2003. IEMDC'03.}, vol.~1, 2003, pp. 11--15 vol.1.

\bibitem{178163}
R.~Krishnan and A.~Bharadwaj, ``A comparative study of various motor drive systems for aircraft applications,'' in \emph{Conference Record of the 1991 IEEE Industry Applications Society Annual Meeting}, 1991, pp. 252--258 vol.1.

\bibitem{7098414}
B.~Sarlioglu and C.~T. Morris, ``More electric aircraft: Review, challenges, and opportunities for commercial transport aircraft,'' \emph{IEEE Transactions on Transportation Electrification}, vol.~1, no.~1, pp. 54--64, 2015.

\bibitem{4504786}
T.~M. Jahns, G.~B. Kliman, and T.~W. Neumann, ``Interior permanent-magnet synchronous motors for adjustable-speed drives,'' \emph{IEEE Transactions on Industry Applications}, vol. IA-22, no.~4, pp. 738--747, 1986.

\bibitem{6064067}
P.~B. Reddy, A.~El-Refaie, K.-K. Huh, J.~K. Tangudu, and T.~M. Jahns, ``Comparison of interior and surface pm machines equipped with fractional-slot concentrated windings for hybrid traction applications,'' in \emph{2011 IEEE Energy Conversion Congress and Exposition}, 2011, pp. 2252--2259.

\bibitem{7862256}
A.~Tüysüz, F.~Meyer, M.~Steichen, C.~Zwyssig, and J.~W. Kolar, ``Advanced cooling methods for high-speed electrical machines,'' \emph{IEEE Transactions on Industry Applications}, vol.~53, no.~3, pp. 2077--2087, 2017.

\bibitem{9992038}
D.~Lee, T.~Balachandran, S.~Sirimanna, N.~Salk, A.~Yoon, P.~Xiao, J.~Macks, Y.~Yu, S.~Lin, J.~Schuh, P.~Powell, and K.~S. Haran, ``Detailed design and prototyping of a high power density slotless pmsm,'' \emph{IEEE Transactions on Industry Applications}, vol.~59, no.~2, pp. 1719--1727, 2023.

\bibitem{6646754}
A.~M. El-Refaie, J.~P. Alexander, S.~Galioto, P.~Reddy, K.-K. Huh, P.~de~Bock, and X.~Shen, ``Advanced high power-density interior permanent magnet motor for traction applications,'' in \emph{2013 IEEE Energy Conversion Congress and Exposition}, 2013, pp. 581--590.

\bibitem{N35UH}
\BIBentryALTinterwordspacing
\emph{Sintered Neodymium-Iron-Boron Magnets}, Arnold Magnetic Technologies, rev. 210607. [Online]. Available: \url{https://www.arnoldmagnetics.com/wp-content/uploads/2017/11/N35UH-151021.pdf}
\BIBentrySTDinterwordspacing

\bibitem{7917257}
S.~Li, B.~Sarlioglu, S.~Jurkovic, N.~R. Patel, and P.~Savagian, ``Analysis of temperature effects on performance of interior permanent magnet machines for high variable temperature applications,'' \emph{IEEE Transactions on Industry Applications}, vol.~53, no.~5, pp. 4923--4933, 2017.

\bibitem{7880626}
------, ``Comparative analysis of torque compensation control algorithms of interior permanent magnet machines for automotive applications considering the effects of temperature variation,'' \emph{IEEE Transactions on Transportation Electrification}, vol.~3, no.~3, pp. 668--681, 2017.

\bibitem{7873348}
S.~Li, D.~Han, and B.~Sarlioglu, ``Modeling of interior permanent magnet machine considering saturation, cross coupling, spatial harmonics, and temperature effects,'' \emph{IEEE Transactions on Transportation Electrification}, vol.~3, no.~3, pp. 682--693, 2017.

\bibitem{9628538}
I.~Voncilă, I.~Paraschiv, and M.~Costin, ``The influence of saturation on the performance of pmsm,'' in \emph{2021 7th International Symposium on Electrical and Electronics Engineering (ISEEE)}, 2021, pp. 1--6.

\bibitem{370284}
T.~Sebastian, ``Temperature effects on torque production and efficiency of pm motors using ndfeb magnets,'' \emph{IEEE Transactions on Industry Applications}, vol.~31, no.~2, pp. 353--357, 1995.

\bibitem{4347951}
Y.-S. Kim and S.-K. Sul, ``Torque control strategy of an ipmsm considering the flux variation of the permanent magnet,'' pp. 1301--1307, 2007.

\bibitem{9176}
P.~Pillay and R.~Krishnan, ``Modeling of permanent magnet motor drives,'' \emph{IEEE Transactions on Industrial Electronics}, vol.~35, no.~4, pp. 537--541, 1988.

\bibitem{25541}
------, ``Modeling, simulation, and analysis of permanent-magnet motor drives. i. the permanent-magnet synchronous motor drive,'' \emph{IEEE Transactions on Industry Applications}, vol.~25, no.~2, pp. 265--273, 1989.

\bibitem{1064466}
T.~Sebastian, G.~Slemon, and M.~Rahman, ``Modelling of permanent magnet synchronous motors,'' \emph{IEEE Transactions on Magnetics}, vol.~22, no.~5, pp. 1069--1071, 1986.

\bibitem{43253}
F.~Parasiliti and P.~Poffet, ``A model for saturation effects in high-field permanent magnet synchronous motors,'' \emph{IEEE Transactions on Energy Conversion}, vol.~4, no.~3, pp. 487--494, 1989.

\bibitem{5994848}
S.~L. Kellner and B.~Piepenbreier, ``General pmsm d,q-model using optimized interpolated absolute and differential inductance surfaces,'' in \emph{2011 IEEE International Electric Machines \& Drives Conference (IEMDC)}, 2011, pp. 212--217.

\bibitem{7490387}
G.~Luo, R.~Zhang, Z.~Chen, W.~Tu, S.~Zhang, and R.~Kennel, ``A novel nonlinear modeling method for permanent-magnet synchronous motors,'' \emph{IEEE Transactions on Industrial Electronics}, vol.~63, no.~10, pp. 6490--6498, 2016.

\bibitem{111}
X.~Sun and X.~Xiao, ``Precise non-linear flux linkage model for permanent magnet synchronous motors based on current injection and bivariate function approximation,'' \emph{IET Electric Power Applications}, vol.~14, no.~11, pp. 2044--2050.

\bibitem{en12050783}
\BIBentryALTinterwordspacing
K.~Drobnič, L.~Gašparin, and R.~Fišer, ``Fast and accurate model of interior permanent-magnet machine for dynamic characterization,'' \emph{Energies}, vol.~12, no.~5, 2019. [Online]. Available: \url{https://www.mdpi.com/1996-1073/12/5/783}
\BIBentrySTDinterwordspacing

\bibitem{6408244}
C.~Choi, W.~Lee, S.-O. Kwon, and J.-P. Hong, ``Experimental estimation of inductance for interior permanent magnet synchronous machine considering temperature distribution,'' \emph{IEEE Transactions on Magnetics}, vol.~49, no.~6, pp. 2990--2996, 2013.

\bibitem{1413524}
K.~Rahman and S.~Hiti, ``Identification of machine parameters of a synchronous motor,'' \emph{IEEE Transactions on Industry Applications}, vol.~41, no.~2, pp. 557--565, 2005.

\bibitem{1233584}
B.~Stumberger, G.~Stumberger, D.~Dolinar, A.~Hamler, and M.~Trlep, ``Evaluation of saturation and cross-magnetization effects in interior permanent-magnet synchronous motor,'' \emph{IEEE Transactions on Industry Applications}, vol.~39, no.~5, pp. 1264--1271, 2003.

\bibitem{732255}
N.~Bianchi and S.~Bolognani, ``Magnetic models of saturated interior permanent magnet motors based on finite element analysis,'' in \emph{Conference Record of 1998 IEEE Industry Applications Conference. Thirty-Third IAS Annual Meeting (Cat. No.98CH36242)}, vol.~1, 1998, pp. 27--34 vol.1.

\bibitem{7051237}
D.~Hu, Y.~M. Alsmadi, and L.~Xu, ``High-fidelity nonlinear ipm modeling based on measured stator winding flux linkage,'' \emph{IEEE Transactions on Industry Applications}, vol.~51, no.~4, pp. 3012--3019, 2015.

\bibitem{6349974}
J.~G. Cintron-Rivera, A.~S. Babel, E.~E. Montalvo-Ortiz, S.~N. Foster, and E.~G. Strangas, ``A simplified characterization method including saturation effects for permanent magnet machines,'' in \emph{2012 XXth International Conference on Electrical Machines}, 2012, pp. 837--843.

\bibitem{5994804}
T.~Herold, D.~Franck, E.~Lange, and K.~Hameyer, ``Extension of a d-q model of a permanent magnet excited synchronous machine by including saturation, cross-coupling and slotting effects,'' in \emph{2011 IEEE International Electric Machines \& Drives Conference (IEMDC)}, 2011, pp. 1363--1367.

\bibitem{4629410}
K.~J. Meessen, P.~Thelin, J.~Soulard, and E.~A. Lomonova, ``Inductance calculations of permanent-magnet synchronous machines including flux change and self- and cross-saturations,'' \emph{IEEE Transactions on Magnetics}, vol.~44, no.~10, pp. 2324--2331, 2008.

\bibitem{7001597}
X.~Chen, J.~Wang, B.~Sen, P.~Lazari, and T.~Sun, ``A high-fidelity and computationally efficient model for interior permanent-magnet machines considering the magnetic saturation, spatial harmonics, and iron loss effect,'' \emph{IEEE Transactions on Industrial Electronics}, vol.~62, no.~7, pp. 4044--4055, 2015.

\bibitem{7258372}
X.~Chen, J.~Wang, and A.~Griffo, ``A high-fidelity and computationally efficient electrothermally coupled model for interior permanent-magnet machines in electric vehicle traction applications,'' \emph{IEEE Transactions on Transportation Electrification}, vol.~1, no.~4, pp. 336--347, 2015.

\bibitem{7409111}
S.~Lin, X.~Li, T.~Wu, L.~Chow, Z.~Tang, and S.~Stanton, ``Temperature dependent reduced order ipm motor model based on finite element analysis,'' in \emph{2015 IEEE International Electric Machines \& Drives Conference (IEMDC)}, 2015, pp. 543--549.

\bibitem{4676996}
F.~Poltschak and W.~Amrhein, ``A dynamic nonlinear model for permanent magnet synchronous machines,'' in \emph{2008 IEEE International Symposium on Industrial Electronics}, 2008, pp. 724--729.

\bibitem{8557671}
H.~Cai and D.~Hu, ``On pmsm model fidelity and its implementation in simulation,'' in \emph{2018 IEEE Energy Conversion Congress and Exposition (ECCE)}, 2018, pp. 1674--1681.

\bibitem{18878}
T.~Sebastian and G.~Slemon, ``Transient modeling and performance of variable-speed permanent-magnet motors,'' \emph{IEEE Transactions on Industry Applications}, vol.~25, no.~1, pp. 101--106, 1989.

\bibitem{Dehkordi2005PermanentMS}
\BIBentryALTinterwordspacing
A.~B. Dehkordi, A.~Gole, and T.~Maguire, ``Permanent magnet synchronous machine model for real-time simulation,'' 2005. [Online]. Available: \url{https://api.semanticscholar.org/CorpusID:15953835}
\BIBentrySTDinterwordspacing

\bibitem{fitzgerald}
A.~E. Fitzgerald, C.~Kingsley, and S.~D. Umans, \emph{Electric Machinery}.\hskip 1em plus 0.5em minus 0.4em\relax McGraw-Hill, 2014.

\bibitem{en14092343}
\BIBentryALTinterwordspacing
M.~Gierczynski and L.~M. Grzesiak, ``Comparative analysis of the steady-state model including non-linear flux linkage surfaces and the simplified linearized model when applied to a highly-saturated permanent magnet synchronous machine—evaluation based on the example of the bmw i3 traction motor,'' \emph{Energies}, vol.~14, no.~9, 2021. [Online]. Available: \url{https://www.mdpi.com/1996-1073/14/9/2343}
\BIBentrySTDinterwordspacing

\bibitem{9923908}
P.~Alvarez, M.~Satrústegui, I.~Elósegui, and M.~Martinez-Iturralde, ``Review of high power and high voltage electric motors for single-aisle regional aircraft,'' \emph{IEEE Access}, vol.~10, pp. 112\,989--113\,004, 2022.

\bibitem{Armando2013ExperimentalIO}
\BIBentryALTinterwordspacing
E.~Armando, R.~I. Bojoi, P.~Guglielmi, G.~Pellegrino, and M.~Pastorelli, ``Experimental identification of the magnetic model of synchronous machines,'' \emph{IEEE Transactions on Industry Applications}, vol.~49, pp. 2116--2125, 2013. [Online]. Available: \url{https://api.semanticscholar.org/CorpusID:13836038}
\BIBentrySTDinterwordspacing

\bibitem{6939721}
G.~Pellegrino, B.~Boazzo, and T.~M. Jahns, ``Magnetic model self-identification for pm synchronous machine drives,'' \emph{IEEE Transactions on Industry Applications}, vol.~51, no.~3, pp. 2246--2254, 2015.

\end{thebibliography}

\begin{IEEEbiographynophoto}{Kishan Srinivasan}
 received the B.Tech degree in electrical engineering from Pandit Deendayal Energy University, India in 2019 and MS degree from the University of Michigan, Ann Arbor in 2021, where he is currently pursuing the Ph.D. degree with the Department of Electrical Engineering and Computer Science. His research interests include power electronics, electric machines and drives.
\end{IEEEbiographynophoto}
\begin{IEEEbiographynophoto}{Heath F. Hofmann}
(S’90–M’92–SM’16-F'22) received the Ph.D. degree in electrical engineering and computer science from the University of California at Berkeley, Berkeley, CA, USA, in 1998. He is currently a Professor with the University of Michigan, Ann Arbor, MI, USA. He has authored approximately five dozen papers in refereed journals, and currently holds 8 patents. His research interests include power electronics, specializing in the design, simulation, and control of electromechanical systems.
\end{IEEEbiographynophoto}
\begin{IEEEbiographynophoto}{Jing Sun}
received the Ph.D. degree from the University of Southern California at Los Angeles, Los Angeles, CA, USA, in 1989. She is currently the Michael G. Parsons Collegiate Professor with the Department of Naval Architecture and Marine Engineering, with joint appointments with the Department of Electrical Engineering and Computer Science, and the Department of Mechanical Engineering, University of Michigan, Ann Arbor, MI, USA. Her research interests include modeling, control, and optimization of dynamic systems, with applications to marine and automotive systems.
Dr. Sun is a fellow of the National Academy of Inventors, International Federation of Automatic Control (IFAC), and the Society of Naval Architects and Marine Engineers. She was a recipient of the 2003 IEEE Control System Technology Award.
\end{IEEEbiographynophoto}

\end{document}